\documentclass[superscriptaddress,aps,prd,nofootinbib,twocolumn,showpacs]{revtex4-1}
\usepackage{amsfonts}
\usepackage{amsmath}
\usepackage{amssymb}
\usepackage{graphicx}
\usepackage{mathrsfs}
\usepackage{hhline,color}
\usepackage{pifont,ulem}
\usepackage{lmodern}
\usepackage{epsfig}
\usepackage{enumitem} 
\usepackage{ulem}
\usepackage[utf8]{inputenc}
\usepackage{multirow}
\usepackage[abs]{overpic}
\usepackage{pst-plot}
\usepackage{multido}
\definecolor{mypink}{RGB}{219, 48, 222}
\definecolor{brown}{RGB}{200,150, 50}

\newcommand{\pc}[1]{\ensuremath{\left(#1\right)}}

\def\beq{\begin{equation}}
\def\eeq{\end{equation}}
\def\beqa{\begin{eqnarray}}
\def\eeqa{\end{eqnarray}}
\def\ban{\begin{eqnarray*}}
\def\ean{\end{eqnarray*}}
\def\bi{\begin{itemize}}
\def\ei{\end{itemize}}

\newcommand{\Z}{\mathbb{Z}}

\begin{document}

\title{Net baryon-number fluctuations in magnetized quark matter}

\author{Márcio Ferreira}
\email{mferreira@teor.fis.uc.pt}
\affiliation{CFisUC, Department of Physics,
University of Coimbra, P-3004 - 516  Coimbra, Portugal}
\author{Pedro Costa}
\email{pcosta@uc.pt}
\affiliation{CFisUC, Department of Physics,
University of Coimbra, P-3004 - 516  Coimbra, Portugal}
\author{Constança Providência}
\email{cp@fis.uc.pt}
\affiliation{CFisUC, Department of Physics,
University of Coimbra, P-3004 - 516  Coimbra, Portugal}

\date{\today}

\begin{abstract}
The kurtosis and skewness of net baryon-number fluctuations are studied for the 
magnetized phase diagram of three-flavor quark matter within the Polyakov 
extended Nambu--Jona-Lasinio model. 
Two models with magnetic catalysis and inverse magnetic catalysis are considered.
Special attention is given to their behavior in the neighborhood of the light 
and strange critical end points (CEPs). 
Several isentropic trajectories that come close the CEPs are studied in order 
to analyze possible signatures of a CEP in the presence of external magnetic 
fields. 
The effect of the magnetic field on the velocity of sound, $v_s^2$, when both 
the light and strange CEPs are approached from the crossover region is also 
investigated by calculating their temperature and baryon chemical potential 
dependencies at fixed distances from these CEPs. 
Regions with large fluctuations but no CEP in nonmagnetized matter develop a 
CEP under the action of a strong magnetic field.
Besides, the Landau quantization of the quark trajectories may result in the 
appearance of extra CEPs, in particular, in the strange sector for strong
magnetic fields, identifiable by the net baryon-number fluctuations. 
Stiffer (smoother) fluctuations in the region of the CEP are characteristic of 
models that do not predict (do predict) the inverse magnetic catalysis at zero 
chemical potential.
Particularly interesting is the ratio $\chi^4_B/\chi^2_B$ that has a more 
pronounced peak structure, indicating that it is eventually a more convenient 
probe for the search of a CEP.
The speed of sound shows a much richer structure in magnetized quark matter and
allows one to identify both chiral and deconfinement transitions.
\end{abstract}

\maketitle

\section{Introduction}

Notable theoretical and experimental efforts \cite{Brambilla:2014jmp} are being 
done to uncover the rich details of the QCD phase structure 
\cite{Halasz:1998qr,Gupta:2011wh}, namely the nature of the hadron matter-quark 
gluon plasma phase transition and the eventual existence of the QCD chiral 
critical endpoint (CEP) in the phase diagram.

Experimentally, one of the main goals of the heavy ion collision (HIC) programs
has been to unveil the possible existence and location of the CEP on the QCD 
phase diagram. This topic has experienced  great developments over the last few 
years \cite{Abelev:2009bw,Aggarwal:2010cw,Aduszkiewicz:2015jna}. 
Nevertheless, the location of the CEP is still a mystery, its search being a 
major goal of several ongoing and future HIC experiments; the search for the 
CEP is being undertaken in Super Proton Synchrotron (SPS) (NA61/SHINE 
Collaboration) at CERN \cite{Grebieszkow:2017gqx,Davis:2017mzd};  in the 
Relativistic Heavy Ion Collider (RHIC) (STAR Collaboration) at Brookhaven 
National Laboratory \cite{Adamczyk:2014fia,Adamczyk:2017wsl}; and in future 
facilities Facility for Antiproton and Ion Research (FAIR) at GSI 
Helmholtzzentrum für Schwerionenforschung \cite{Ablyazimov:2017guv}, 
Nuclotron-based Ion Collider fAcility (NICA) at Joint Institute for Nuclear 
Research \cite{NICAWP}, J-PARC Heavy Ion Project at Japan Proton Accelerator 
Research Complex (J-PARC) \cite{Sako:J-PARC} (a review on the experimental 
search of the CEP can be found in Ref. \cite{Akiba:2015jwa}).
In addition, both NICA and J-PARC-HI HIC programs are expected to create
extremely dense matter comparable to the neutron star core, where the eventual 
first-order phase boundary can also be explored.

Therefore, it is important to probe the QCD phase transition and the possible 
existence of the CEP by investigating potential measurable signatures that phase 
transitions can leave in the final state of HIC experiments (see Ref. 
\cite{Senger:2011zza}).

Fluctuations of conserved quantities, such as baryon, electric charge, and the 
strangeness number, are very important to the experimental search for the 
CEP in relativistic HIC. Indeed, measurements of cumulants of the net proton 
(proxy for net baryon) \cite{Adamczyk:2013dal}, net charge 
\cite{Adamczyk:2014fia}, and net kaon (proxy for net strangeness) 
\cite{Adamczyk:2017wsl} are expected to provide relevant information on the 
medium created by the collision (for a review, see Refs. 
\cite{Friman:2011pf,Asakawa:2015ybt,Braun-Munzinger:2015hba,Luo:2017faz}).
Experimentally, these quantities are studied by measuring event-by-event 
fluctuations: a given observable is measured on an event-by-event basis, and its 
fluctuations are studied for the ensemble of events 
\cite{Braun-Munzinger:2015hba}.

Due to the second-order nature of the phase transition that occurs at the CEP, 
divergences of correlation lengths for a static system of infinite size will 
take place.
Therefore, cumulants of the net baryon number diverge 
\cite{Stephanov:1998dy,Stephanov:1999zu}, making them particularly interesting. 
Kurtosis \cite{Stephanov:2011pb} and skewness \cite{Asakawa:2009aj}
for the net baryon-number fluctuation distributions are connected to high-order 
cumulants, which can be extracted from event-by-event fluctuations in HIC
experiments. Moreover, once they consist of cumulants ratios they are 
independent of the volume of the system.

By using the (2+1)-flavor Nambu--Jona-Lasinio (NJL) model, the study of 
fluctuations of conserved charges (baryon number, electric charge, 
and strangeness) at finite temperature and density has been done in Refs.
\cite{Luo:2017faz,Chen:2015dra,Fan:2016ovc}. 
The study of these fluctuations employing the (2+1)-flavor
Polyakov--Nambu--Jona-Lasinio (PNJL) \cite{Costa:2010zw} model was performed 
in Refs. \cite{Fu:2009wy,Fu:2010ay,Bhattacharyya:2010ef,Shao:2017yzv} at finite 
temperature and in Refs. \cite{Fu:2010ay,Shao:2017yzv,Liu:2017qyc} at finite 
temperature and density.

Several conditions can affect the eventual existence and location of the CEP. 
It is known that the presence of external magnetic fields is one of 
them\footnote{A highly inhomogeneous magnetic field with a value of about 
5$m_{\pi}^2$ is formed in some HIC, even if for a very short time 
\cite{Voronyuk:2011jd}.}.
Other circumstances that affect the location of the CEP are the strangeness and 
the isospin content of the medium 
\cite{Costa:2013zca,Costa:2015bza,Rechenberger:2016gim}. 
Also considering the effects of repulsive vector interactions and of 
the inverse magnetic catalysis (IMC)  mechanism 
\cite{Ferreira:2013tba,Ferreira:2014kpa}, which have opposite competing 
effects, will dramatically influence the position of the CEP in the phase 
diagram \cite{Costa:2015bza}.

The effect of magnetic fields is especially fascinating and a very timely topic
\cite{Fukushima:2012kc,Miransky:2015ava,Andersen:2014xxa}. 
Several low-energy effective models, including the NJL-type models, have 
been used to investigate the impact of strong magnetic fields at finite 
temperature 
\cite{Ferreira:2013oda,Ferreira:2014a,Farias:2014eca,Ayala:2014gwa,Ayala:2016bbi,Pagura:2016pwr} and at finite baryonic chemical potentials 
\cite{Avancini:2012ee,Chao:2013qpa,Ayala:2015lta}.
Indeed, it was found that external magnetic fields induce several CEPs
in the strange sector \cite{Ferreira:2017wtx}, which arise due to the multiple 
phase transitions that the strange quark undergoes. The same happens for the 
light sector by taking isospin breaking chemical potential
\cite{Costa:2013zca,Rechenberger:2016gim}. 
Moreover, other regions of the QCD phase diagram are affected by magnetic 
fields like the first phases of the Universe 
\cite{Vachaspati:1991nm,Enqvist:1993np} and compact stellar objects 
\cite{Menezes:2014aka}.

It becomes crucial, therefore, to understand how an external magnetic field
affects the structure of the QCD phase diagram, namely its impact on the 
fluctuations of net baryon number. At finite temperature, this was done in Ref. 
\cite{Fu:2013ica}. We now will study this effect at finite temperature, $T$, and 
baryonic chemical potential, $\mu_B$, giving attention to both the light and 
strange sectors, with the respective CEPs and associated first-order phase 
transitions.

In this work, we study the magnetized phase diagram using the (2+1)-flavor PNJL 
model from the point of view of the kurtosis and skewness of net baryon-number 
fluctuations near the light and strange CEPs, in both the crossover and 
first-order transition regions.
The model and formalism are presented in Sec. \ref{sec:model}, while the results 
are in Sec. \ref{sec:Results}.
Finally, in Sec. \ref{Conclusions}, we draw our conclusions. 

\section{Model and Formalism}
\label{sec:model}

\subsection{PNJL model}

The magnetized three-flavor quark matter is investigated using the (2+1)-flavor
Nambu--Jona-Lasinio model coupled to the Polyakov loop. 
The model Lagrangian density reads
\begin{eqnarray}
{\cal L} &=& {\bar{q}} \left[i\gamma_\mu D^{\mu}-
	{\hat m}_c \right ] q ~+~ {\cal L}_\text{sym}~+~{\cal L}_\text{det}~-~\mathcal{U}\left(\Phi,\bar\Phi;T\right)\nonumber\\
& &~+~\frac{1}{4}F_{\mu \nu}F^{\mu \nu}.
	\label{Pnjl}
\end{eqnarray}
The ${\cal L}_\text{sym}$ and ${\cal L}_\text{det}$ denote, respectively, 
the scalar-pseudoscalar interaction and the 't Hooft six-fermion interaction
\cite{Klevansky:1992qe,Hatsuda:1994pi},
\begin{align}
	{\cal L}_\text{sym}&= G_s \sum_{a=0}^8 \left [({\bar q} \lambda_ a q)^2 + 
	({\bar q} i\gamma_5 \lambda_a q)^2 \right ] \\
	{\cal L}_\text{det}&=-K\left\{{\rm det} \left [{\bar q}(1+\gamma_5)q \right] + 
	{\rm det}\left [{\bar q}(1-\gamma_5)q\right] \right \}.
\end{align}
The (eletro)magnetic tensor is given by 
$F^{\mu \nu }=\partial^{\mu }A_{EM}^{\nu }-\partial ^{\nu }A_{EM}^{\mu }$,
with $A_{EM}^\mu$ being the external (electro)magnetic field.
We consider a  static and constant magnetic field in the $z$ direction, 
$A_{EM}^\mu=\delta^{\mu 2} x_1 B$.
The quark field is represented in flavor space by $q = (u,d,s)^T$, with 
(current) mass matrix ${\hat m}_c= {\rm diag}_f (m_u,m_d,m_s)$. 
The Gell-Mann matrices are denoted by $\lambda_a$. 
The external (electro)magnetic field couples with both the quarks and the 
effective gluon field, $A^\mu$, through the covariant derivative,
$D^{\mu}=\partial^\mu - i q_f A_{EM}^{\mu}-i A^\mu$\footnote{The quark electric 
charges are $q_d = q_s = -q_u/2 = -e/3$, where $e$ 
is the electron charge.}. 
The effective gluon field is given by 
$A^\mu = g_{strong} {\cal A}^\mu_a\frac{\lambda_a}{2}$, where
${\cal A}^\mu_a$ represents the SU$_c(3)$ gauge field.
The spatial components are neglected in Polyakov gauge at finite temperature,
i.e., $A^\mu = \delta^{\mu}_{0}A^0 = - i \delta^{\mu}_{4}A^4$. 
The Polyakov loop value is defined as the trace of the Polyakov line,
$ \Phi = \frac 1 {N_c} {\langle\langle \mathcal{P}\exp i\int_{0}^{\beta}d\tau\,
A_4\left(\vec{x},\tau\right)\ \rangle\rangle}_\beta$,
which is the order parameter of the $\Z_3$ symmetric/broken phase transition in 
pure gauge. For the pure gauge sector, we use the following effective potential 
\cite{Roessner:2006xn},
\begin{eqnarray}
	& &\frac{\mathcal{U}\left(\Phi,\bar\Phi;T\right)}{T^4}
	= -\frac{a\left(T\right)}{2}\bar\Phi \Phi \nonumber\\
	& &
	+\, b(T)\mbox{ln}\left[1-6\bar\Phi \Phi+4(\bar\Phi^3+ \Phi^3)
	-3(\bar\Phi \Phi)^2\right],
	\label{Ueff}
\end{eqnarray}
where 
$a\left(T\right)=a_0+a_1\left(\frac{T_0}{T}\right)+a_2\left(\frac{T_0}{T}\right)^2$ and $b(T)=b_3\left(\frac{T_0}{T}\right)^3$. 
Its parametrization values are: 
$a_0 = 3.51$, $a_1 = -2.47$, $a_2 = 15.2$, and $b_3 = -1.75$ 
\cite{Roessner:2006xn}, while the critical temperature is set to $T_0=210$ MeV 
in order to reproduce the pseudocritical temperature for the deconfinement 
coming from lattice calculations \cite{Aoki:2009sc}.

The divergent ultraviolet sea quark integrals are regularized by a sharp cutoff 
$\Lambda$ in three-momentum space.
For the NJL model parametrization, we consider $\Lambda = 602.3$ MeV, 
$m_u= m_d=5.5$ MeV, $m_s=140.7$ MeV, $G_s^0 \Lambda^2= 1.835$, and 
$K \Lambda^5=12.36$ \cite{Rehberg:1995kh}.

At finite magnetic field, two model variants with distinct scalar couplings are 
considered: the usual NJL model with constant $G_s=G_s^0$ coupling and a 
magnetic field dependent coupling $G_s=G_s(eB)$ \cite{Ferreira:2014kpa}.
The latter model gives a decrease of both the chiral and deconfinement 
pseudocritical temperatures at $\mu_B=0$ with increasing magnetic field
strength, in accordance with lattice QCD (LQCD) calculations 
\cite{baliJHEP2012}, while the opposite occurs for the $G_s^0$ model. Its functional dependence is 
$G_s(\zeta)=G_s^0\pc{\frac{1+a\,\zeta^2+b\,\zeta^3}{1+c\,\zeta^2+d\,\zeta^4}}$,
where $\zeta=eB/\Lambda_{QCD}^2$ (with $\Lambda_{QCD}=300$ MeV).
The parameters are $a = 0.0108805$, $b=-1.0133\times10^{-4}$, $c= 0.02228$, and 
$d=1.84558\times10^{-4}$ \cite{Ferreira:2014kpa}.
Both models coincide at zero magnetic field, $G_s=G_s^0=G_s(eB=0)$.
The gap equations are solved within the mean field approximation (the equations 
can be found in Refs. \cite{Menezes:2008qt,Menezes:2009uc}). 

\subsection{Baryon-number susceptibilities}

\begin{figure*}[!t]
	\centering
	\includegraphics[width=1.0\linewidth]{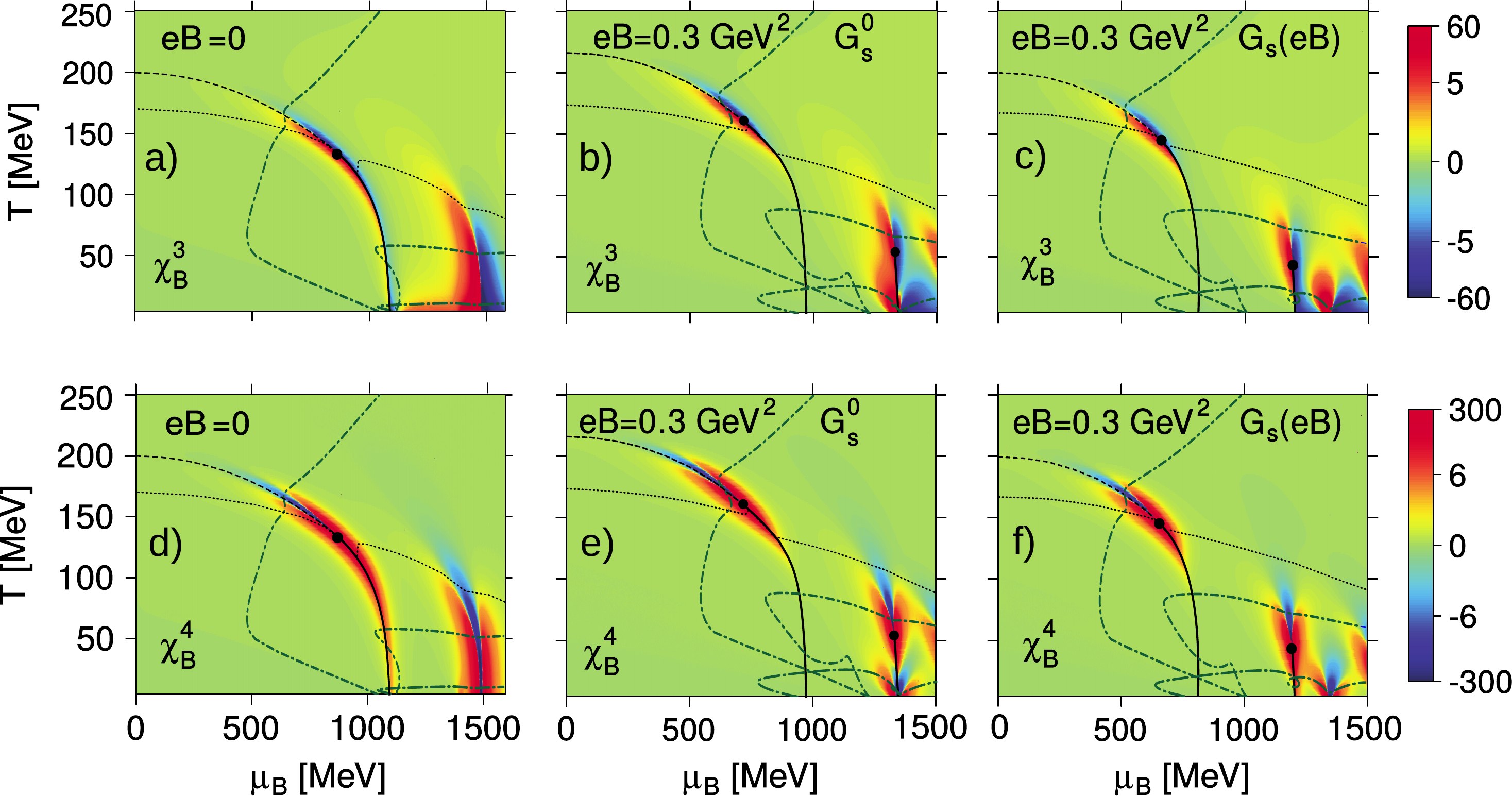}
	\caption{The $\chi^3_B$ (top panels) and $\chi^4_B$ (bottom panels) 
	net baryon-number susceptibilities for zero magnetic field [panels a) and d)]
	and $eB=0.3$ GeV$^2$ within the $G_s^0$ [panels b) and e)] and $G_s(eB)$ 
	[panels c) and f)] models. The following information is shown: the CEP 
	(black dot), the chiral first-order phase transition (black solid 
	line), both the chiral (black dashed line) and deconfinement (black dotted
	line) crossovers, and the $s/\rho_B=15,1,0.1$ isentropes (dark green 
	solid-dashed lines), which appear in the clockwise direction, respectively.
  } 
	\label{fig:1}
\end{figure*}
Fluctuations or cumulants of conserved charges, such as baryon number, provide
crucial information on critical phenomena.
In a thermal equilibrium medium, fluctuations of conserved charges behave 
characteristically, enabling the identification of the onset of deconfinement 
or the possible existence of a CEP on the QCD phase diagram. 
They are then expected to provide a characteristic signature for the presence 
of a CEP that can be experimentally observed.
Herein, we focus on the baryon-number charge fluctuations. 
The n${th}$-order net baryon (generalized) susceptibility is given by
\begin{eqnarray}
\chi_B^n(T,\mu_B)= \frac{\partial^n\pc{P(T,\mu_B)/T^4}}{\partial(\mu_B/T)^n}\,.
\end{eqnarray}
Due to the extensivity property of cumulants, different susceptibilities ratios,
\begin{eqnarray}
\chi_B^{n,m}\equiv\frac{\chi_B^n(T,\mu_B)}{\chi_B^m(T,\mu_B)},
\end{eqnarray}
are then calculated in order to eliminate the volume dependence, allowing for a
possible comparison with experimental observables. 
In this work, we analyze the following ratios,
\begin{align}
&\chi_B^{4,2}(T,\mu_B)=\frac{\chi_B^4(T,\mu_B)}{\chi_B^2(T,\mu_B)}=\kappa\sigma^2,\\
&\chi_B^{3,1}(T,\mu_B)=\frac{\chi_B^3(T,\mu_B)}{\chi_B^1(T,\mu_B)}=\frac{S_B\sigma^3}{M},
\end{align}
where $M=VT^3\chi_B^1$ is the mean, $\sigma^2=VT^3\chi_B^2$ is the variance, 
$S_B$ is the skewness, and $\kappa$ is the kurtosis of the net baryon-number 
distribution. 

\section{Results}
\label{sec:Results}

Herein, we analyze the qualitative dependence of the net baryon-number 
fluctuations over the phase diagram and the effect of an external magnetic 
field. We consider quark matter with equal quark chemical potentials, 
$\mu_u=\mu_d=\mu_s=\mu_q$. 
The baryonic chemical potential is then given by $\mu_B=3\mu_q$.
For a finite magnetic field, we perform a comparison between the $G_s=G_s^0$ and 
$G_s=G_s(eB)$ models (both models coincide at zero magnetic field).

The $\chi_B^3$ and $\chi_B^4$ net baryon-number susceptibilities for $eB=0$ 
(left panels) and $0.3$ GeV$^2$ (middle panels for the $G_s^0$ model and right 
panels for the $G_s(eB)$ model) are given in Fig. \ref{fig:1}.
For a better understanding  of their dependence on $T$ and $\mu_B$, we present 
the following information on the plots: 
the chiral (dashed black line) and the deconfinement (dotted black line) 
pseudocritical boundaries, the first-order chiral phase transition (black 
solid line), CEP (black dot), and three isentropic trajectories (dark green 
dashed lines), i.e., trajectories along which the system entropy over the 
baryon density, $s/\rho_B$, is constant. 
The pseudocritical boundaries, where the crossover transition is characterized 
by an analytic behavior, allow for different definitions of (pseudo)critical 
temperature through different observables.
The pseudocritical temperature is often defined as the temperature at which the 
susceptibility of the order parameters takes its maximum value (the point where 
fluctuations are largest). Using this definition, the PNJL model presents 
$T_{\chi}^{ps}(\mu_B=0)=200$ MeV and $T_{\Phi}^{ps}(\mu_B=0)=171$ MeV at $B=0$.
However, LQCD results presents a different order for these crossovers:
$T_{\chi}=157$ \cite{Borsanyi:2010bp} and $T_{\Phi}^{ps}(\mu_B=0)=170$ 
\cite{Aoki:2009sc}. 
In general, in the PNJL type models (or the Polyakov-Quark-Meson model), the 
temperature of the transitions is inverted.
In the PNJL model we have two sectors to determine the respective scales: 
the NJL one (fitted to chiral symmetry breaking phenomenology in vacuum, a scale 
related to the strength of the condensate) and the gluonic one (fixed by pure 
gauge results at finite temperature with the scale $T_0$).
The coupling is done via the covariant derivative. If $T_0$ is allowed to vary, 
it is possible to control the relative scales of the transitions and to obtain 
the correct scale hierarchy for the transitions by increasing  $T_0$.
However, in this case, both the chiral and the deconfinement transitions occur 
at too high temperatures \cite{Hubert_p_c}.

The following qualitative features remain valid in the presence or absence of 
$B$ and, therefore, are valid for the three scenarios of Fig. \ref{fig:1}:
(i) the $\chi_B^3$ values are asymmetric with respect to the chiral transitions,
with the $\chi_B^3>0$ region on the broken chiral symmetry region, and
(ii) the $\chi_B^4>0$ region is nearly symmetric with respect to the 
$\chi_B^4<0$ one, which lies along the chiral crossover boundary.

Regardless of the absence of a first-order phase transition for the strange 
quark at $eB=0$, a region with a nonmonotonic dependence, very similar to the 
one around the CEP, is seen for high values of $\mu_B$ (see panels a) and d) of 
Fig. \ref{fig:1}).  
This result, which is attributed to the specific model parametrization 
employed \cite{Ferreira:2018sun}, is signaling the proximity of a first-order 
phase transition for the strange quark; a stronger scalar coupling would 
eventually give rise to a first-order phase transition also for this sector. 
In fact, a first-order phase transition for the strange sector automatically 
appears when eight-quark interactions are included \cite{Moreira:2014qna}. 

The change of the chiral transition from crossover to first order in strong 
magnetic fields is related with the filling of Landau levels, and therefore, at 
least an additional CEP emerges at high $\mu_B$ and at low $T$ (the second black 
dot in middle and right panels of Fig. \ref{fig:1}) for both scalar coupling 
models (see Ref. \cite{Ferreira:2017wtx}). 
The appearance of multiple CEPs in the strange sector was already reported in 
Ref. \cite{Ferreira:2017wtx}, where the analysis of the strange quark condensate 
(with $T_0=270$ MeV) showed multiple phase transitions.
Furthermore, multiple phase transitions for the light quarks also occur when 
lower $eB$ values are considered \cite{Costa:2013zca}.
The $G_s(eB)$ model predicts both CEPs at lower $\mu_B$ and $T$ values than the 
$G_s^0$ model. 

\begin{figure*}[!t]
	\centering
	\includegraphics[width=1.0\linewidth]{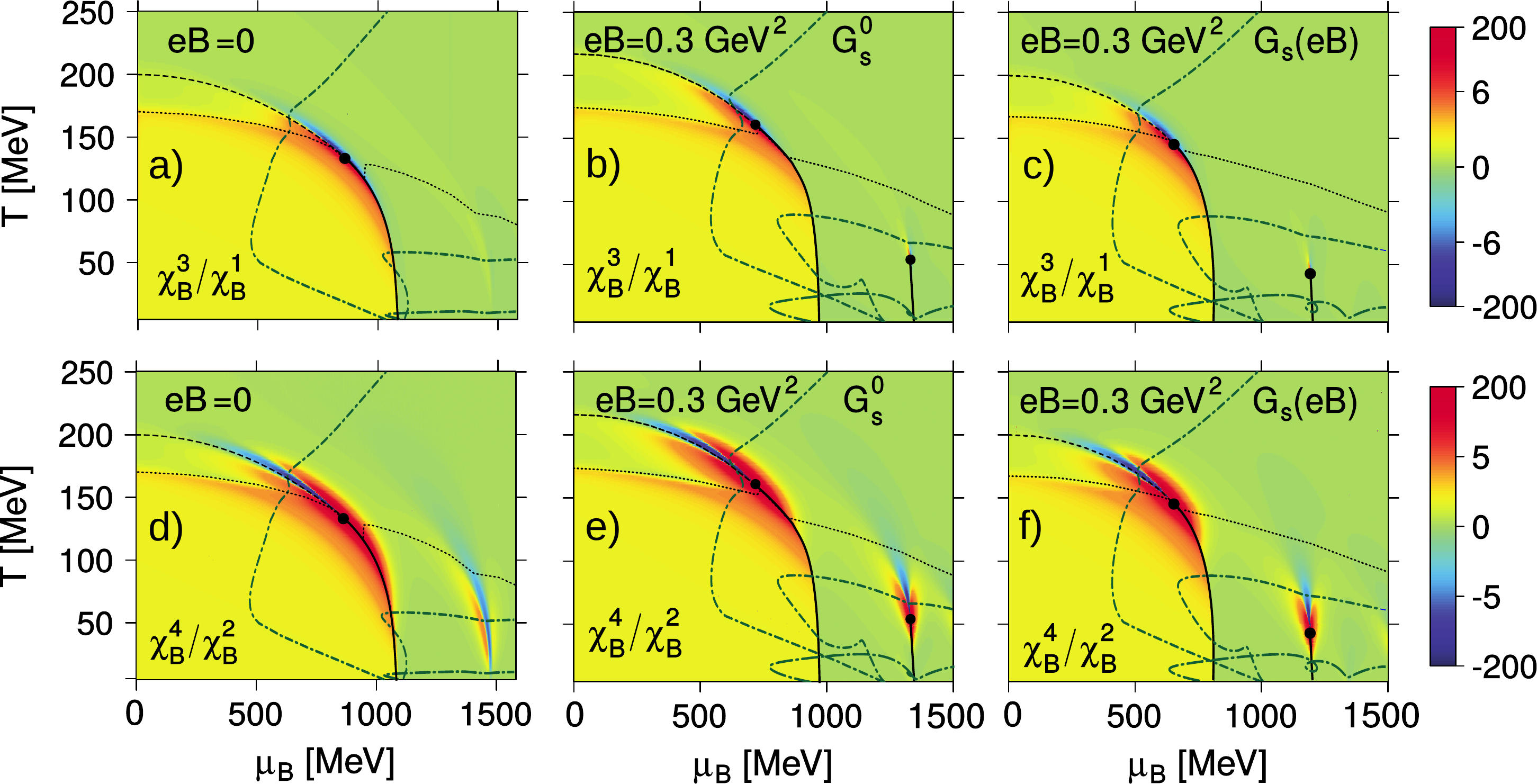}
	\caption{The $\chi^3_B/\chi^1_B$ (top panels) and $\chi^4_B/\chi^2_B$ 
	(bottom panels) net baryon-number susceptibilities for zero magnetic field 
	[panels a) and)] and $eB=0.3$ GeV$^2$ within the $G_s^0$ [panels b) and e)]
	and $G_s(eB)$ [panels c) and f)] models. The following information is shown: 
	the critical point (black dot), the chiral first-order phase transition 
	(black solid line), both the chiral (black dashed line) and deconfinement 
	(black dotted line) crossovers, and the $s/\rho_B=15,1,0.1$ 
	isentropes (dark green solid-dashed	lines), which appear in the clockwise
	direction, respectively.} 
	\label{fig:2}
\end{figure*}

It is interesting that two additional nonmonotonic susceptibility regions are 
present, indicating critical regions, at around $\mu_B\approx1350$ MeV (see 
panels b) and especially e) of Fig. \ref{fig:1} for $\chi_B^4$): a critical 
region at higher temperatures, associated to the strange CEP, and a second one 
at low temperatures with no connection with a first-order phase transition. 
This second critical region is due to a fast increase of the density of 
$d$ quarks at small temperatures.
When the $G_s(eB)$ model is considered, these two critical regions occur at  
different values of $\mu_B$ (see panels c) and f)): the critical region related 
with the strange quark and the respective first-order transition with the CEP 
are pushed to lower baryonic chemical potentials, while the critical region 
associated with the $d$ quark is practically unchanged because its mass is 
already close to the current mass. However, even in the absence of a CEP, this 
critical region looks like a ``near-CEP'' region. 
At even higher values of $\mu_B$, a glimpse of a new critical region appears.
This is the second first-order phase transition for the strange quark found in 
Ref. \cite{Ferreira:2017wtx}.

To examine possible signatures of a CEP in nearby isentropes, we have 
determined three specific isentropic trajectories (dark green solid-dashed 
lines in all panels of Fig. \ref{fig:1}): one that passes above the light CEP 
(i.e., in the crossover region), $s/\rho_B=15$, which will be analyzed in detail 
later; and two that take place in the low $T$ and high $\mu_B$ region, 
$s/\rho_B=1$ and $0.1$ (clockwise direction).

At high baryonic chemical potentials, the $s/\rho_B=1$ isentrope shows a 
characteristic behavior (bending towards the CEP as seen in panels b) and c) or
e) and f)) near the nonmonotonic susceptibility regions for the strange CEP;
due to the low temperature of these isentropes, a sudden decrease/increase of 
the strange quark density must be balanced by a rapid decrease/increase of the 
temperature to keep $s/\rho_B$ constant.
Additionally, the $s/\rho_B=0.1$ isentrope for the $G_s(eB)$ model shows a 
bending toward the negative (blue) region of the $\chi_B^4$ value at 
$\mu_B\approx1350$ MeV (see Fig. \ref{fig:1}, panel f)). 
This is the already referred critical region without the presence of a CEP.
Around the light first-order phase transition both isentropes behave as seen in
Ref. \cite{Ferreira:2017wtx}.

The ratios $\chi^3_B/\chi^1_B$ and $\chi^4_B/\chi^2_B$ are plotted in Fig. 
\ref{fig:2}. 
Aside from the region near the CEP of the light quarks, 
$\chi^3_B/\chi^1_B$ also shows the presence of a slight nonmonotonic behavior 
near the CEP related with the strange quark phase transition, especially in the 
presence of the magnetic field (see panels a), b), and c) of Fig. \ref{fig:2}). 
The $\chi^4_B/\chi^2_B$, otherwise, shows a well-defined nonmonotonic 
structure near both CEPs (see panels d), e) and f) of Fig. \ref{fig:2}) but no 
signal of the multiple structure seen in Fig. \ref{fig:1} for $\chi^3_B$ and 
$\chi^4_B$. 
Both ratios show a clear distinction between the regions where chiral symmetry 
is broken and (approximately) restored. 
They are also sensitive to the deconfinement transition as shown by their 
pronounced variation along the deconfinement boundary at lower baryonic 
chemical potentials (dotted black line).
Even if both ratios are sensitive to the presence of external magnetic fields, 
$\chi^4_B/\chi^2_B$ has a more pronounced peak structure, which indicates that 
it is probably a more useful probe for the CEP and for the strong magnetic 
field produced in early noncentral collisions.
Figure \ref{fig:2} also shows an interesting difference of the magnetic fields 
effect on chiral and deconfinement transitions when the $G_s^0$ model or the 
$G_s(eB)$ model is considered. Indeed, as already pointed out, the PNJL model 
presents $T_{\chi}^{ps}(\mu_B=0)=200$ MeV and $T_{\Phi}^{ps}(\mu_B=0)=171$ 
MeV at $B=0$, and thus a gap of $29$ MeV between the pseudocritical transition 
temperatures. At finite $B$, the gap increases for the $G_s^0$ model, 
whereas it remains almost unchanged for the $G_s(eB)$ model.

\begin{figure*}[!htbp]
	\centering
	\includegraphics[width=0.8\linewidth]{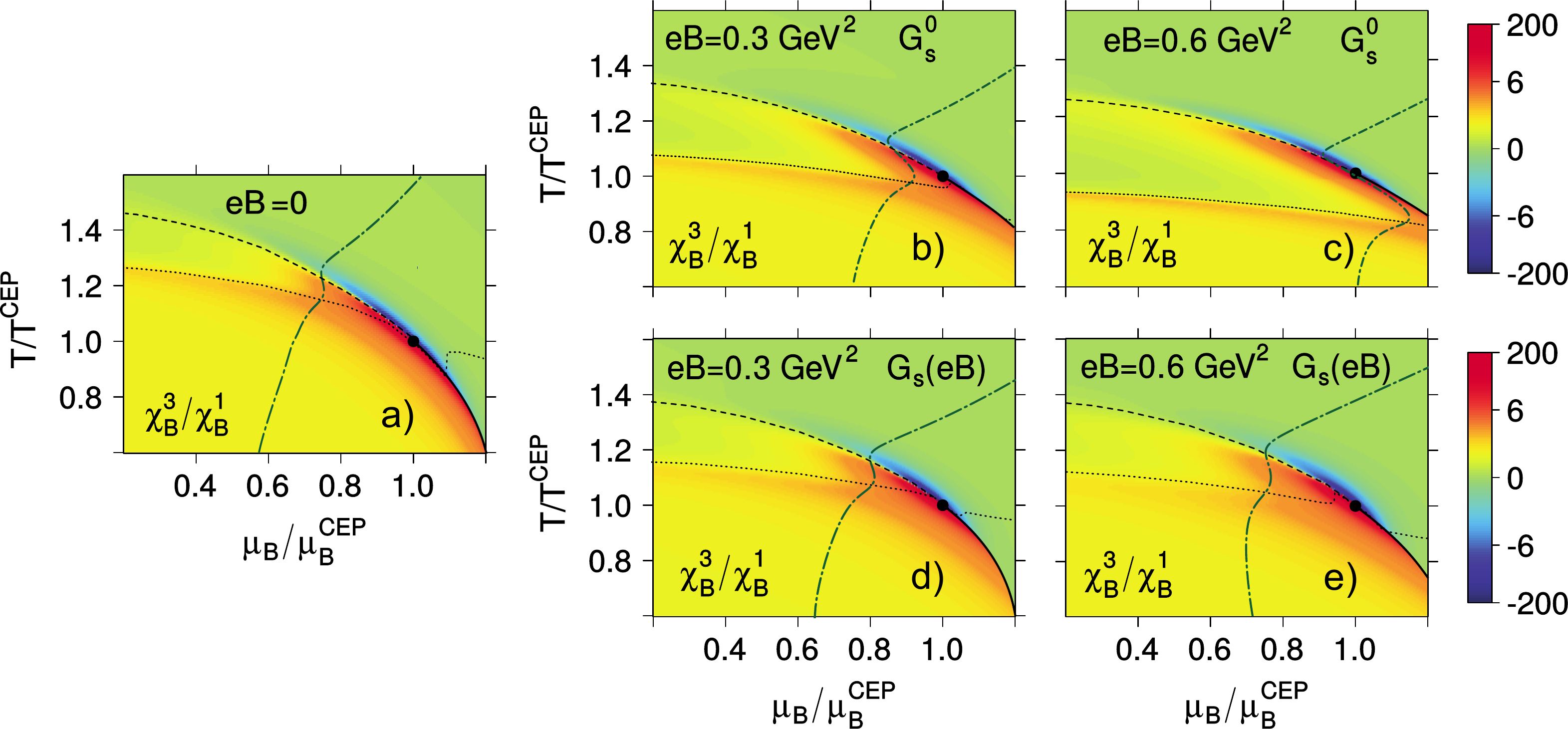}
	\caption{The $\chi^3_B/\chi^1_B$ net baryon-number susceptibility
	around the chiral CEP for $eB=0$ [panel a)], $eB=0.3$ GeV$^2$ within the 
	$G_s^0$ [panel b)] and $G_s(eB)$ [panel d)] models and $eB=0.6$ GeV$^2$
	within the $G_s^0$ [panel c)] and $G_s(eB)$ [panel e)] models.
	The following information is shown: the CEP (black dot), 	the chiral 
	first-order phase transition (black solid line), both the chiral 
	(black dashed line) and deconfinement (black dotted line) crossovers, and the
	$s/\rho_B=15$ isentrope (dark green solid-dashed line).
	} 
	\label{fig:3}
\end{figure*}
\begin{figure*}[!htbp]
	\centering
	\includegraphics[width=0.8\linewidth]{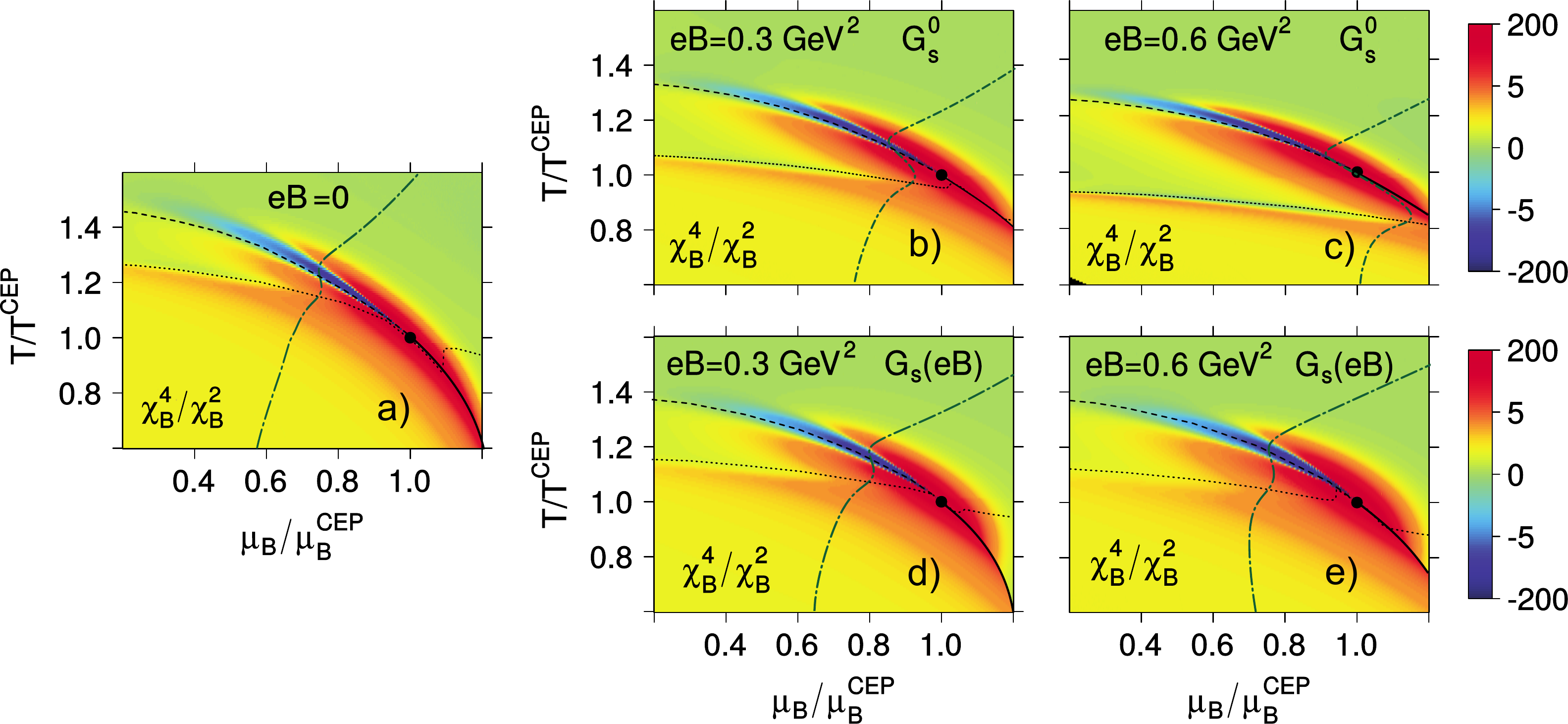}
	\caption{The $\chi^4_B/\chi^2_B$ net baryon-number susceptibility
	around the chiral CEP for $eB=0$ [panel a)], $eB=0.3$ GeV$^2$ within the 
	$G_s^0$ [panel b)] and $G_s(eB)$ [panel d)] models and $eB=0.6$ GeV$^2$
	within the $G_s^0$ [panel c)] and $G_s(eB)$ [panel e)] models.
	The following information is shown: the CEP (black dot),
	the chiral first-order phase transition (black solid line), both the chiral 
	(black dashed line) and deconfinement (black dotted line) crossovers, and the
	$s/\rho_B=15$ isentrope (dark green solid-dashed line).
	}  
	\label{fig:4}
\end{figure*}

Let us now focus on the dependence of $\chi^3_B/\chi^1_B$ and 
$\chi^4_B/\chi^2_B$ around the CEP of the light quarks.
They are plotted as a function of $\mu_B/\mu_B^{\texttt{CEP}}$ and  
$T/T^{\texttt{CEP}}$ in, respectively, Figs. \ref{fig:3} and \ref{fig:4},  
at $eB=0,\,0.3$ and $0.6$ GeV$^2$ for both models.  
The isentrope $s/\rho_B=15$ (dark green solid-dashed line) is also shown.
The phase diagram in terms of the ratios $\mu_B/\mu_B^{\texttt{CEP}}$ and
$T/T^{\texttt{CEP}}$ turns visible some features of the behavior of the light 
quark condensate due to the magnetic field independently of the location of 
the CEP. Indeed, the enhancement of both fluctuation ratios at low $\mu_B$ due 
to $B$ reflects the effect of the magnetic field on the CEP location, 
$(T^{\texttt{CEP}},\mu_B^{\texttt{CEP}})$, and the respective extension of the 
critical region.
Let us first consider the $G_s^0$ model: the increase of the magnetic field 
localizes the strong fluctuations closer to CEP and to the deconfinement 
crossover and separates the chiral symmetry restoration and deconfinement 
crossovers with a valleylike feature. This is seen for both ratios. 
The $G_s(eB)$ model behaves differently: the high fluctuation range spreads to 
a larger extension when $B$ increases, and the region between the two 
crossover lines is filled with larger fluctuation ratios. 
To summarize, the $G_s(eB)$ model, which describes the IMC effect, gives rise to 
smoother fluctuations that are detected in a larger region of the phase diagram.
This is also visible in Fig. \ref{fig:3}, where it is not seen a considerable 
enhancing of the fluctuation region, i.e., the size of the nonmonotonic region 
seems to be independent of the $B$ strength for each model.
The $G_s(eB)$ model shows an enlargement of the nonmonotonic region of the 
fluctuation ratio with increasing magnetic field when compared with the $G_s^0$
model.

Finally, the $G_s(eB)$ model predicts that $\mu_B^{\texttt{CEP}}$ decreases 
with $B$, and the chiral crossover at $\mu_B=0$ possibly turns into a 
first-order phase transition for high enough $B$ \cite{Costa:2015bza}. 
This would lead to an increase of the fluctuation ratio at low values of $\mu_B$ 
due to the dragging of the critical region by the CEP.
Although the $G_s^0$ model also predicts the same tendency for $eB<0.3$ 
GeV$^2$, $\mu_B^{\texttt{CEP}}$ increases for higher values of $B$ and 
consequently, fluctuations now are reduced at low values of $\mu_B$.
A strong enhancement of $\chi_B^{3(4)}/\chi_B^{1(2)}$ at $\mu_B=0$ with 
increasing $B$ would, therefore, signals the vicinity of a CEP at small $\mu_B$ 
values.

The $s/\rho_B=15$ isentropic trajectory at intermediate temperatures and 
chemical potentials shows some similarities for all panels of Fig. \ref{fig:3} 
(Fig. \ref{fig:4} presents the same isentropic line).
We see that a change in the isentropic trajectory occurs while crossing the 
chiral (black dashed line) and the deconfinement (black dotted line) phase 
boundaries: as the temperature decreases, a first bend in the $s/\rho_B=15$ 
isentrope occurs just slightly above the chiral crossover, and a second bend in 
the opposite direction occurs just slightly above the deconfinement transition.
The chemical potential and temperature decrease continuously along the 
isentrope, except close to the crossing of the crossovers when a more or less 
intense backbending of the line to higher chemical potential occurs. The 
larger effect occurs for the $G_s^0$ model.

\begin{figure}[!t]
	\centering
	\includegraphics[width=0.8\linewidth]{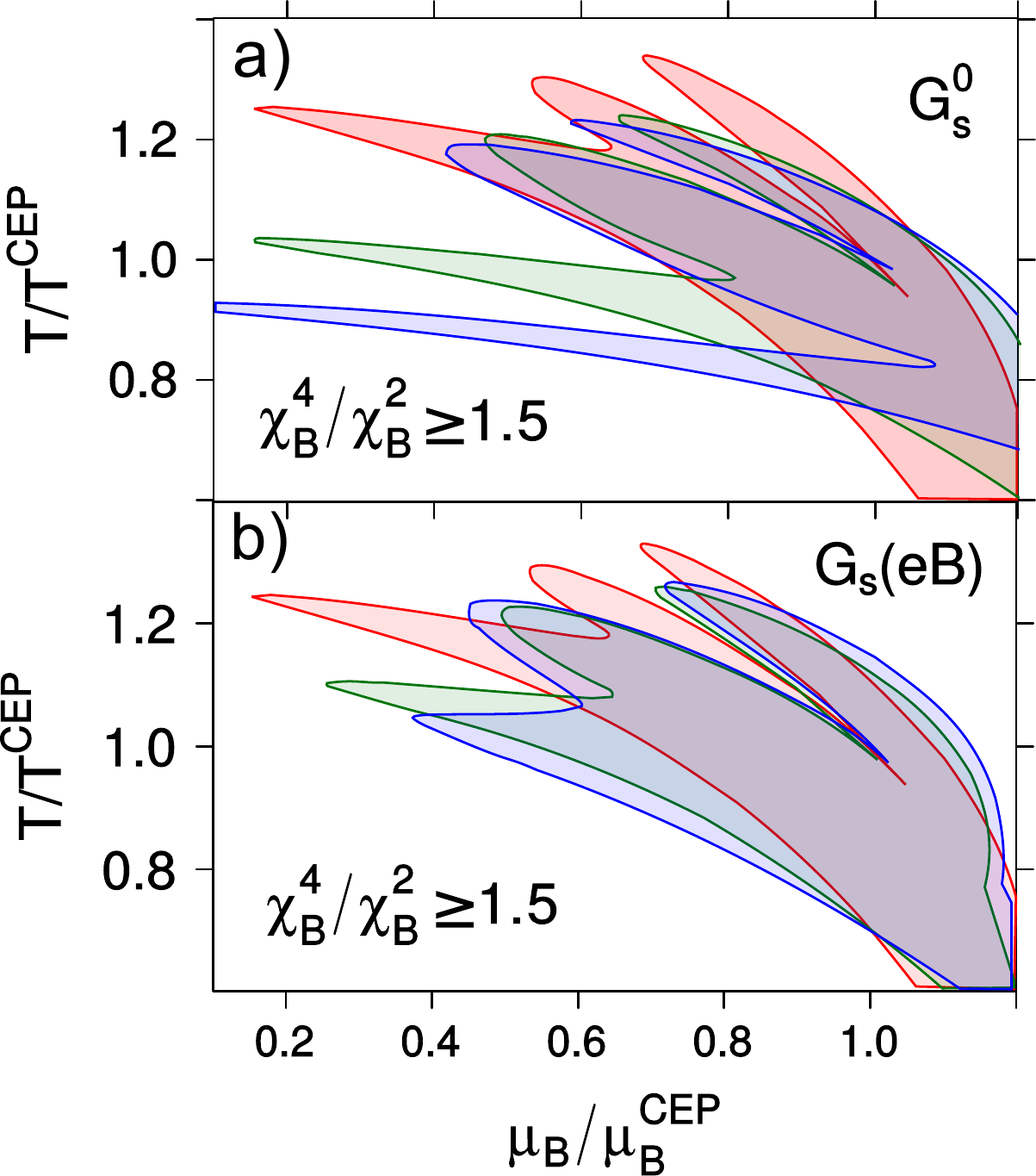}
	\caption{Region around the chiral CEP with $\chi_B^4/\chi_B^2\geq1.5$ for 
	$eB=0$ GeV$^2$ (red), $0.3$ GeV$^2$ (green), and $0.6$ GeV$^2$ (blue) within
	the $G_s^0$ (top) and $G_s(eB)$ (bottom) models.
	} 
	\label{fig:5}
\end{figure}
\begin{figure*}[!t]
	\centering
	\includegraphics[width=0.8\linewidth]{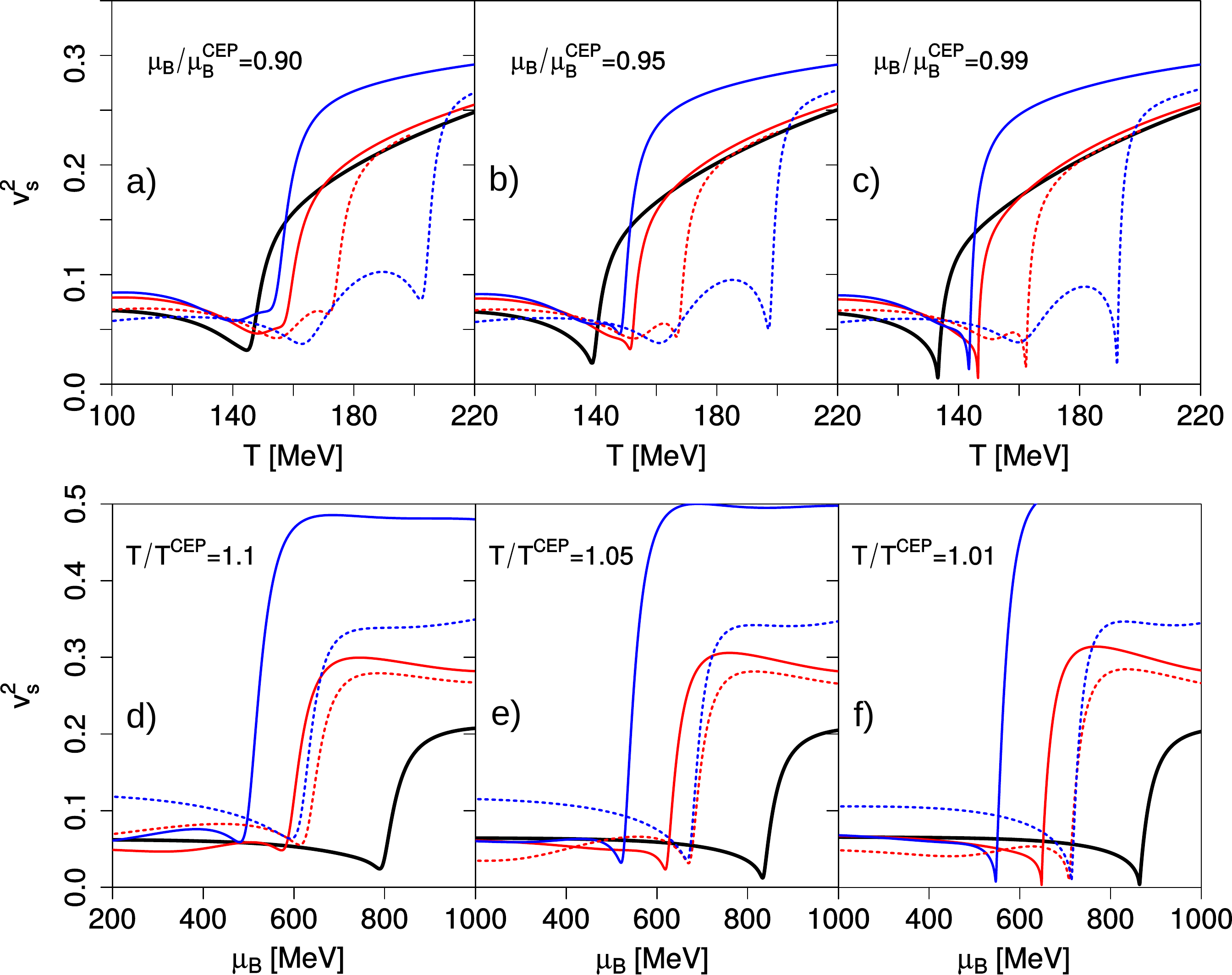}
	\caption{Sound velocity squared, $v_s^2$, as a function of temperature (top 
	panels) and baryonic chemical potential (bottom panels) for $eB=0$ (black 
	lines), $0.3$ GeV$^2$ (red lines), and $0.6$ GeV$^2$ (blue lines) within the 
	$G_s^0$ (dashed) and $G_s(eB)$ (solid) models around the light CEP. 
	Three values for the relative distance to CEP are considered: the $T$ 
	dependence is computed at $\mu_B/\mu_B^{\texttt{CEP}}=0.90$ [panel a)], $0.95$ 
	[panel b)], and $0.99$ [panel c)], whereas the $\mu_B$ dependence is
	determined at $T/T^{\texttt{CEP}}=1.1$ [panel d)], $1.05$ [panel e)], 
	and $1.01$ [panel f)].
	} 
	\label{fig:6}
\end{figure*}

In order to investigate how the strong fluctuation region is affected by the 
magnetic field, we show the regions where $\chi^4_B/\chi^2_B\geq1.5$ in Fig. 
\ref{fig:5}.
The results for $eB=0$ (red region), $0.3$ (green region), and $0.6$ (blue 
region) GeV$^2$ for the $G_s^0$ (panel a)) and $G_s(eB)$ (panel b)) models 
are presented and show that the area of the $\chi^4_B/\chi^2_B\geq1.5$ regions 
does not depend significantly on the magnetic field.
The main effect for the $G_s^0$ model is the rotation of the whole region and 
the separation of the third brunch (at lower temperatures), due to the growing 
gap between the chiral and the deconfinement transition with increasing $B$. 
The $G_s(eB)$ model, however, only shows a small rotation and a rather 
constant gap between the crossover transitions. The rotation of the whole 
region is also expected since the $T^{\texttt{CEP}}$ has a smaller increase for 
the $G_s(eB)$ model than for the $G_s^0$ model. 
As referred to before, another effect is the widening of the bands along or 
between the crossover lines within the $G_s(eB)$ model, corresponding to a 
smoother behavior of the condensate.

\begin{figure*}[!t]
	\centering
	\includegraphics[width=0.8\linewidth]{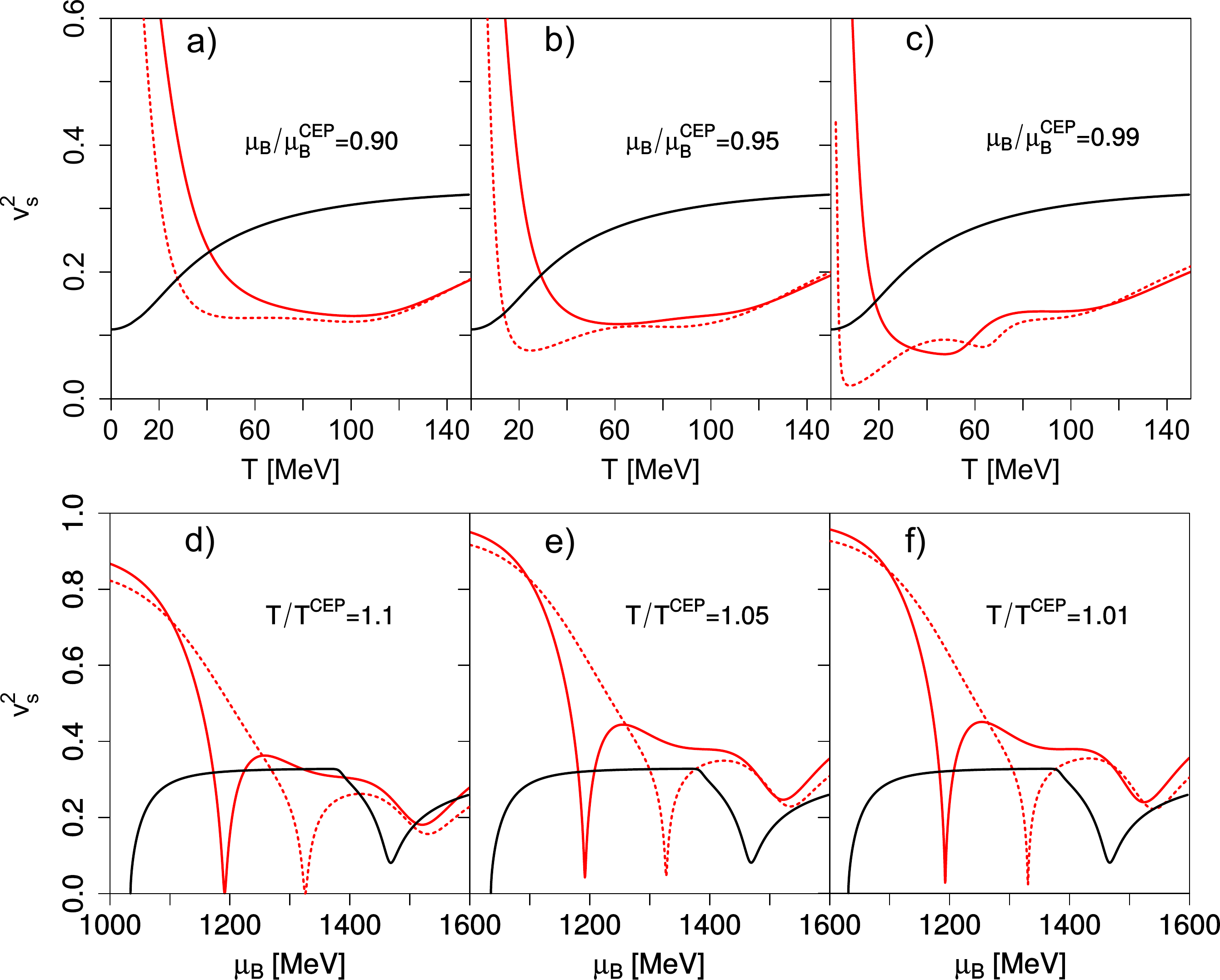}
	\caption{Sound velocity squared, $v_s^2$, as a function of temperature (top 
	panels) and baryonic chemical potential (bottom panels) for $eB=0.3$ GeV$^2$
	within the $G_s^0$ (dashed) and $G_s(eB)$ (solid lines) models around the
	strange CEP. Three values for the relative distance to CEP are considered: 
	the $T$ 	dependence is computed at $\mu_B/\mu_B^{\texttt{CEP}}=0.90$ [panel 
	a)], $0.95$ [panel b)], and $0.99$ [panel c)], whereas the $\mu_B$ dependence
	is determined at $T/T^{\texttt{CEP}}=1.1$ [panel d)], $1.05$ [panel e)], 
	and $1.01$ [panel f)]. 
	The black line for all panels is the sound velocity squared at $eB=0$ in the
	near-CEP region for the strange sector: in panels a), b) and c), $v_s^2(T)$
	is calculated at $\mu_B\approx1473$ MeV; and in panels e), f) and g), 
	$v_s^2(\mu_B)$ is calculated at $T=1$ MeV.
	}
	\label{fig:7}
\end{figure*}

The sound velocity, $v_s^2=\partial P/\partial {\cal E}$, is a fundamental 
quantity in the expansion of hot and dense matter \cite{Rafelski:2015cxa}. 
To investigate the effect of the magnetic field on $v_s^2$ when the light CEP
is approached from the crossover region, we calculate its $T$ and $\mu_B$ 
dependencies at fixed distances from the CEP.
The temperature dependence, $v_s^2(T)$, is determined at 
$\mu_B/\mu_B^{\texttt{CEP}}=\{0.90,0.95,0.99\}$,
while the $\mu_B$ dependence, $v_s^2(\mu_B)$, is calculated at 
$T/T^{\texttt{CEP}}=\{1.1,1.05,1.01\}$.
We show in Fig. \ref{fig:6} the results for $eB=0$ (black), $0.3$ GeV$^2$ (red), 
and $0.6$ GeV$^2$ (blue) within the $G_s^0$ (dashed lines) and $G_s(eB)$ (solid 
lines) models.

While the $B=0$ results show only a local minimum that tends to zero in an 
increasingly stiffer way as the CEP gets closer (independently if we approach 
the CEP by temperature or baryonic chemical potentials as seen in both top and 
bottom panels of Fig. \ref{fig:6}, black line), two local minimum, related with 
the deconfinement and chiral crossovers transitions, are present in both models 
for finite $B$ (see the red and blue curves in Fig. \ref{fig:6}). 
Indeed, at $B=0$, when we are very close to  the CEP in the crossover region, 
both transitions coincide (see for example panel a) of Fig. \ref{fig:4}): 
with the parametrization we are using, we have different pseudocritical 
temperatures at $\mu_B=0$ ($T_{\chi}^{ps}=200$ MeV and $T_{\Phi}^{ps}=171$ MeV) 
for $eB=0$, but this difference almost vanishes at 
$\mu_B/\mu_B^{\texttt{CEP}}=0.90$, giving rise to just one minimum in the sound 
velocity (black line).
At $\mu_B/\mu_B^{\texttt{CEP}}=0.90$, in the presence of a magnetic field, the 
lowest minimum is due to the deconfinement transition, and the second minimum is 
related to the chiral transition (see the red and blue curves in panel a) of 
Fig. \ref{fig:6}). This second minimum tends to zero as the CEP gets closer 
(see case $\mu_B/\mu_B^{\texttt{CEP}}=0.99$, panel c), in the same figure).
Thus, although the sound propagation in hot matter that passes near the CEP 
slows down in the critical region, far from the CEP, the decrease of sound 
velocity is more pronounced on the deconfinement transition. 
Once again, the increasing gap between the pseudocritical temperatures for the 
$G_s^0$ model with $B$ is clear in the top panels of Fig. \ref{fig:6}: the two 
minima separate from each other (see the dashed lines). 
The effect of both transitions is also visible  in the $v_s^2(\mu_B)$ 
dependence (panels d), e), and f) of the same figure).

Finally, we analyze the effect of the magnetic field on the square of the 
velocity of sound $v_s^2$ when the strange CEP is approached. 
In Fig. \ref{fig:7}, we show results for magnetized matter with $eB=0.3$ 
GeV$^2$, red  dashed(solid) lines corresponding to the $G_s^0$($G_s(eB)$) 
model, and nonmagnetized matter (black line). 
The $T$ and $\mu_B$ dependencies of $v_s^2$ are calculated at fixed distances 
from the CEP: the temperature dependence, $v_s^2(T)$, is determined at 
$\mu_B/\mu_B^{\texttt{CEP}}=\{0.90,0.95,0.99\}$, and the $\mu_B$ dependence, 
$v_s^2(\mu_B)$, is calculated at $T/T^{\texttt{CEP}}=\{1.1,1.05,1.01\}$, and for 
nonmagnetized matter we take $\mu_B\approx1473$ MeV and $T=1$ MeV, both close 
to a near-CEP region for the strange sector.

We first consider the temperature dependence at fixed $\mu_B$.
The $G_s^0$ model shows an interesting feature, not seen for the $G_s(eB)$ 
model, as the CEP is approached in the $\mu_B$ direction from below  (dashed 
lines in panels a), b), and c) of Fig. \ref{fig:7}).
The first minimum of $v_s^2(T)$ at lower values of $T$ corresponds to the 
critical region, but with no CEP, related with the $d$ quark (see panels b) and
d) of Fig. \ref{fig:1}). The second minimum corresponds to the strange first-
order transition.
The $G_s(eB)$ model shows only this last minimum related with the strange 
first-order transition. The black line calculated for $eB=0$ in the 
near-CEP region shows a monotonic behavior, decreasing as $T$ decreases, 
signaling the approach of the near CEP.

We now study the $\mu_B$ dependence of  $v_s^2$ at fixed $T$.
When we approach the CEP from above taking several values of $T$, (panels d), e)
, and f) of Fig. \ref{fig:7}), $v_s^2(\mu_B)$ shows at lower $\mu_B$ a peaked 
minimum in both models of magnetized quark matter. These minima correspond to 
the minimum that the black line for $eB=0$ shows at $\mu_B\approx1473$ MeV, the 
near-CEP region.
For $eB=0$, $v_s^2$ becomes negative at the lowest values of $\mu_B$ shown. 
This signals the entry into the first-order region of the light quarks that 
occurs at a much larger $\mu_B$ than in the magnetized matter considered.
The $G_s(eB)$ model (red-solid lines) shows a second minimum around 
$\mu_B\approx1350$ MeV, due to the critical region with no CEP connected to the 
$d$ quark. 
At high $\mu_B$, another minimum occurs again for both models. This is related 
with a second CEP for the strange quark found in Ref. \cite{Ferreira:2017wtx} 
that appears at lower temperatures. 

\section{Conclusions}
\label{Conclusions}

The effect of the magnetic field on the QCD phase diagram, and, in particular, 
on the kurtosis and the skewness of the net baryon-number fluctuations, is 
discussed. Two different models of the (2+1)-flavor PNJL model are considered: 
one with the usual constant scalar coupling, $G_s^0$, and the other with a 
magnetic field dependent coupling, $G_s(eB)$, including the IMC for the 
chiral symmetry restoration crossover at zero chemical potential.
Both the symmetry chiral restoration of the light and the strange sectors are 
discussed. The kurtosis and the skewness of the net baryon-number fluctuations 
are calculated all over the QCD phase diagram, and a special attention is given 
to their behavior in the neighborhood of the light-quark and strange quark CEP. 
In order to understand which kind of signatures could possibly be identified in 
heavy ion collisions, several isentropic lines that come close the CEP are 
studied. Another property that is analyzed is the velocity of sound in the 
vicinity of both CEPs.

Several conclusions are drawn from the present study. First of all, we have 
confirmed that the presence of the magnetic field may result in the appearance 
of extra CEPs, in particular, in the strange sector. 
For some of the new CEPs, the chemical potential localization of
the CEP is already identifiable in the nonmagnetized QCD phase diagram. This 
is true  for the strange sector. However, there are other critical regions that 
originate from the Landau quantization of the quark trajectories and appear 
only in magnetized matter. 
In the present study, this was observed with the identification of a critical
region connected with the $d$ quark. The identification of the change on the
critical behavior, in particular, of the appearance of new CEPs, was carried 
out from the analysis of the kurtosis and the skewness of the net baryon-number 
fluctuations. In magnetized matter the behavior of $\chi_B^n$ fluctuations up 
to fourth order close to the critical regions is stiffer, and even more stiff 
if the $G_s^0$ model is taken. Considering the fluctuation ratios some of these 
differences disappear but several features could still distinguish the two 
models for magnetized matter. The $G_s(eB)$ model with IMC shows a smoother 
behavior so that the fluctuation ratios spread over a larger region. This was 
particularly true for the region between the two crossover transitions at low 
chemical potential.
Moreover, the $\chi^4_B/\chi^2_B$ has a more pronounced peak structure, which 
indicates that it may be a more convenient probe for the CEP and even for the 
strong magnetic field produced in early noncentral collisions.

The behavior of the sound velocity close the CEPs is reflecting in a very clear 
way the changes in the QCD phase diagram originated by the magnetic field; in 
particular, it is sensitive to both, the deconfinement and the chiral symmetry 
restoration, transitions.
Close to critical regions, this quantity always presents a depression, which may 
have a stiffer behavior in the proximity of a CEP or a  smoother behavior if 
only a region with large fluctuations but no CEP. 
These last regions of phase diagram may, however, transform into CEPs if stronger 
magnetic fields come into play.


\begin{center}
{\bf Acknowledgments:}
\end{center}

We thank Hubert Hansen for several enlightening discussions.
This work was supported by ``Fundação para a Ciência e Tecnologia,'' Portugal, 
under Projects No. UID/FIS/04564/2016, No. POCI-01-0145-FEDER-029912 [with 
financial support from POCI, in its FEDER component, and by the FCT/MCTES 
budget through national funds (OE)] and under Grants No. SFRH/BPD/102273/2014 
(P.C.) and No. CENTRO-01-0145-FEDER-000014 (M.F.) through the CENTRO2020 
program. This work was also partly supported by ``PHAROS,'' COST Action
CA16214, and ``THOR,'' COST Action CA15213.

\end{document}